
\input epsf
%\input youngtab
%%%%%%%%%%%%%%%%%%%%%%%%%%%%%%%%%%%%%%%%%%%%%%%%%%%%%%%%%%%%%%%%%
%                                                               %
%       FONT FAMILIES:                                          %
%                                                               %
%%%%%%%%%%%%%%%%%%%%%%%%%%%%%%%%%%%%%%%%%%%%%%%%%%%%%%%%%%%%%%%%%
%                                                               %
%       Define script letters as rsfs                           %
%               (or redefine as cal)                            %
%                                                               %
%                                                               %
%%%%%%%%%%%%%%%%%%%%%%%%%%%%%%%%%%%%%%%%%%%%%%%%%%%%%%%%%%%%%%%%%
\newfam\scrfam
\batchmode\font\tenscr=rsfs10 \errorstopmode
\ifx\tenscr\nullfont
        \message{rsfs script font not available. Replacing with calligraphic.}
        \def\scr{\cal}
\else   
        \font\sevenscr=rsfs7
        \font\fivescr=rsfs5
        \skewchar\tenscr='177 \skewchar\sevenscr='177 \skewchar\fivescr='177
        \textfont\scrfam=\tenscr \scriptfont\scrfam=\sevenscr
        \scriptscriptfont\scrfam=\fivescr
        \def\scr{\fam\scrfam}
        
\fi
%%%%%%%%%%%%%%%%%%%%%%%%%%%%%%%%%%%%%%%%%%%%%%%%%%%%%%%%%%%%%%%%%
%                                                               %
%       fraktur (or redefine as italic)                     %
%                                                               %
%%%%%%%%%%%%%%%%%%%%%%%%%%%%%%%%%%%%%%%%%%%%%%%%%%%%%%%%%%%%%%%%%
\catcode`\@=11
\newfam\frakfam
\batchmode\font\tenfrak=eufm10 \errorstopmode
\ifx\tenfrak\nullfont
        \message{eufm font not available. Replacing with italic.}
        
\else
    
    \font\sevenfrak=eufm7 \font\fivefrak=eufm5
    \textfont\frakfam=\tenfrak
    \scriptfont\frakfam=\sevenfrak \scriptscriptfont\frakfam=\fivefrak
    
\fi
\catcode`\@=\active
%%%%%%%%%%%%%%%%%%%%%%%%%%%%%%%%%%%%%%%%%%%%%%%%%%%%%%%%%%%%%%%%%
%                                                               %
%       Blackboard bold (or redefine as boldface)               %
%                                                               %
%%%%%%%%%%%%%%%%%%%%%%%%%%%%%%%%%%%%%%%%%%%%%%%%%%%%%%%%%%%%%%%%%
\newfam\msbfam
\batchmode\font\twelvemsb=msbm10 scaled\magstep1 \errorstopmode
\ifx\twelvemsb\nullfont\def\Bbb{\bf}

    \message{Blackboard bold not available. Replacing with boldface.}
\else   \catcode`\@=11
        \font\tenmsb=msbm10 \font\sevenmsb=msbm7 \font\fivemsb=msbm5
        \textfont\msbfam=\tenmsb
        \scriptfont\msbfam=\sevenmsb \scriptscriptfont\msbfam=\fivemsb
        \def\Bbb{\relax\expandafter\Bbb@}
        \def\Bbb@#1{{\Bbb@@{#1}}}
        \def\Bbb@@#1{\fam\msbfam\relax#1}
        \catcode`\@=\active

\fi
%%%%%%%%%%%%%%%%%%%%%%%%%%%%%%%%%%%%%%%%%%%%%%%%%%%%%%%%%%%%%%%%%
%                                                               %
%       More FONTS:                                             %
%                                                               %
%%%%%%%%%%%%%%%%%%%%%%%%%%%%%%%%%%%%%%%%%%%%%%%%%%%%%%%%%%%%%%%%%
        \font\eightrm=cmr8              \def\xrm{\eightrm}
        \font\eightbf=cmbx8             \def\xbf{\eightbf}
        \font\eightit=cmti10 at 8pt     \def\xit{\eightit}
%%%     \font\eightit=cmti8             \def\xit{\eightit}
                     
        \font\eightcp=cmcsc8
        \font\eighti=cmmi8              \def\xold{\eighti}
        \font\eightib=cmmib8             \def\xbold{\eightib}
        \font\teni=cmmi10               \def\old{\teni}
        \font\tencp=cmcsc10

        \font\twelvecp=cmcsc10 scaled\magstep1

     at10pt   
    \font\twelvehelvbold=phvb at12pt
     at14pt
    \font\sixteenhelvbold=phvb at16pt

\def\noblackbox{\overfullrule=0pt}
\noblackbox

%%%%%%%%%%%%%%%%%%%%%%%%%%%%%%%%%%%%%%%%%%%%%%%%%%%%%%%%%%%%%%%%%
%                                                               %
%       HEADLINE:                                               %
%                                                               %
%%%%%%%%%%%%%%%%%%%%%%%%%%%%%%%%%%%%%%%%%%%%%%%%%%%%%%%%%%%%%%%%%
\newtoks\headtext
\headline={\ifnum\pageno=1\hfill\else
    \ifodd\pageno{\eightcp\the\headtext}{ }\dotfill{ }{\old\folio}
    \else{\old\folio}{ }\dotfill{ }{\eightcp\the\headtext}\fi
    \fi}
\def\makeheadline{\vbox to 0pt{\vss\noindent\the\headline\break
\hbox to\hsize{\hfill}}
        \vskip2\baselineskip}
%%%%%%%%%%%%%%%%%%%%%%%%%%%%%%%%%%%%%%%%%%%%%%%%%%%%%%%%%%%%%%%%%
%                                                               %
%       FOOTNOTES:                                              %
%                                                               %
%%%%%%%%%%%%%%%%%%%%%%%%%%%%%%%%%%%%%%%%%%%%%%%%%%%%%%%%%%%%%%%%%
\newcount\infootnote
\infootnote=0
\def\foot#1#2{\infootnote=1
\footnote{${}^{#1}$}{\vtop{\baselineskip=.75\baselineskip
\advance\hsize by -\parindent\noindent{\xrm #2}}}\infootnote=0$\,$}
%%%%%%%%%%%%%%%%%%%%%%%%%%%%%%%%%%%%%%%%%%%%%%%%%%%%%%%%%%%%%%%%%
%                                                               %
%       REFERENCES:                                             %
%                                                               %
%%%%%%%%%%%%%%%%%%%%%%%%%%%%%%%%%%%%%%%%%%%%%%%%%%%%%%%%%%%%%%%%%
\newcount\refcount
\refcount=1
\newwrite\refwrite
\def\oldsize{\ifnum\infootnote=1\xold\else\old\fi}
\def\ref#1#2{
    \def#1{{{\oldsize\the\refcount}}\ifnum\the\refcount=1\immediate\openout\refwrite=\jobname.refs\fi\immediate\write\refwrite{\item{[{\xold\the\refcount}]}
    #2\hfill\par\vskip-2pt}\xdef#1{{\noexpand\oldsize\the\refcount}}\global\advance\refcount by 1}
    }
\def\refout{\catcode`\@=11
        \xrm\immediate\closeout\refwrite
        \vskip2\baselineskip
        {\noindent\twelvecp References}\hfill\vskip\baselineskip
                                                %\vskip.25\baselineskip%%%%
        %\parskip=.875\parskip
        %\baselineskip=.8\baselineskip
        \baselineskip=.75\baselineskip
        \input\jobname.refs
        %\parskip=8\parskip \divide\parskip by 7
        %\baselineskip=1.25\baselineskip
        \baselineskip=4\baselineskip \divide\baselineskip by 3
        \catcode`\@=\active\rm}

\def\hepth#1{\href{http://arxiv.org/abs/hep-th/#1}{hep-th/{\xold#1}}}

\def\jhep#1#2#3#4{\href{http://jhep.sissa.it/stdsearch?paper=#2\%28#3\%29#4}{J. High Energy Phys. {\xbold #1#2} ({\xold#3}) {\xold#4}}}

\def\CMP#1#2#3{Commun. Math. Phys. {\xbold#1} ({\xold#2}) {\xold#3}}
\def\CQG#1#2#3{Class. Quantum Grav. {\xbold#1} ({\xold#2}) {\xold#3}}
\def\IJMPA#1#2#3{Int. J. Mod. Phys. {\xbf A}{\xbold#1} ({\xold#2}) {\xold#3}}
\def\JGP#1#2#3{J. Geom. Phys. {\xbold#1} ({\xold#2}) {\xold#3}}

\def\JMP#1#2#3{J. Math. Phys. {\xbold#1} ({\xold#2}) {\xold#3}}

\def\NPB#1#2#3{Nucl. Phys. {\xbf B}{\xbold#1} ({\xold#2}) {\xold#3}}

\def\PLB#1#2#3{Phys. Lett. {\xbf B}{\xbold#1} ({\xold#2}) {\xold#3}}

\def\PRD#1#2#3{Phys. Rev. {\xbf D}{\xbold#1} ({\xold#2}) {\xold#3}}

%%%%%%%%%%%%%%%%%%%%%%%%%%%%%%%%%%%%%%%%%%%%%%%%%%%%%%%%%%%%%%%%%
%                                                               %
%       SECTION NUMBERING:                                      %
%                                                               %
%%%%%%%%%%%%%%%%%%%%%%%%%%%%%%%%%%%%%%%%%%%%%%%%%%%%%%%%%%%%%%%%%
\newcount\sectioncount
\sectioncount=0
\def\section#1#2{\global\eqcount=0
    \global\subsectioncount=0
        \global\advance\sectioncount by 1
    \ifnum\sectioncount>1
            \vskip2\baselineskip
    \fi
    \noindent
        \line{\twelvecp\the\sectioncount. #2\hfill}
        \vskip.8\baselineskip\noindent
        \xdef#1{{\old\the\sectioncount}}}
\newcount\subsectioncount
\def\subsection#1#2{\global\advance\subsectioncount by 1
    \vskip.8\baselineskip\noindent
    \line{\tencp\the\sectioncount.\the\subsectioncount. #2\hfill}
    \vskip.5\baselineskip\noindent
    \xdef#1{{\old\the\sectioncount}.{\old\the\subsectioncount}}}
\newcount\appendixcount
\appendixcount=0
\def\appendix#1{\global\eqcount=0
        \global\advance\appendixcount by 1
        \vskip2\baselineskip\noindent
        \ifnum\the\appendixcount=1
        \hbox{\twelvecp Appendix: #1\hfill}\vskip\baselineskip\noindent\fi
    \ifnum\the\appendixcount=2
        \hbox{\twelvecp Appendix B: #1\hfill}\vskip\baselineskip\noindent\fi
    \ifnum\the\appendixcount=3
        \hbox{\twelvecp Appendix C: #1\hfill}\vskip\baselineskip\noindent\fi}

%%%%%%%%%%%%%%%%%%%%%%%%%%%%%%%%%%%%%%%%%%%%%%%%%%%%%%%%%%%%%%%%%
%                                                               %
%       EQUATION NUMBERING                                      %
%                                                               %
%%%%%%%%%%%%%%%%%%%%%%%%%%%%%%%%%%%%%%%%%%%%%%%%%%%%%%%%%%%%%%%%%
\newcount\eqcount
\eqcount=0
\def\Eqn#1{\global\advance\eqcount by 1
\ifnum\the\sectioncount=0
    \xdef#1{{\old\the\eqcount}}
    \eqno({\oldstyle\the\eqcount})
\else
        \ifnum\the\appendixcount=0
            \xdef#1{{\old\the\sectioncount}.{\old\the\eqcount}}
                \eqno({\oldstyle\the\sectioncount}.{\oldstyle\the\eqcount})\fi
        \ifnum\the\appendixcount=1
            \xdef#1{{\oldstyle A}.{\old\the\eqcount}}
                \eqno({\oldstyle A}.{\oldstyle\the\eqcount})\fi
        \ifnum\the\appendixcount=2
            \xdef#1{{\oldstyle B}.{\old\the\eqcount}}
                \eqno({\oldstyle B}.{\oldstyle\the\eqcount})\fi
        \ifnum\the\appendixcount=3
            \xdef#1{{\oldstyle C}.{\old\the\eqcount}}
                \eqno({\oldstyle C}.{\oldstyle\the\eqcount})\fi
\fi}
\def\eqn{\global\advance\eqcount by 1
\ifnum\the\sectioncount=0
    \eqno({\oldstyle\the\eqcount})
\else
        \ifnum\the\appendixcount=0
                \eqno({\oldstyle\the\sectioncount}.{\oldstyle\the\eqcount})\fi
        \ifnum\the\appendixcount=1
                \eqno({\oldstyle A}.{\oldstyle\the\eqcount})\fi
        \ifnum\the\appendixcount=2
                \eqno({\oldstyle B}.{\oldstyle\the\eqcount})\fi
        \ifnum\the\appendixcount=3
                \eqno({\oldstyle C}.{\oldstyle\the\eqcount})\fi
\fi}
\def\multi{\global\advance\eqcount by 1}
\def\multieq#1#2{\xdef#1{{\old\the\eqcount#2}}
        \eqno{({\oldstyle\the\eqcount#2})}}
%%%%%%%%%%%%%%%%%%%%%%%%%%%%%%%%%%%%%%%%%%%%%%%%%%%%%%%%%%%%%%%%%
%                                                               %
%       Hyperrefs:                                          %
%                                                               %
%%%%%%%%%%%%%%%%%%%%%%%%%%%%%%%%%%%%%%%%%%%%%%%%%%%%%%%%%%%%%%%%%
\newtoks\url
\def\Href#1#2{\catcode`\#=12\url={#1}\catcode`\#=\active#2}
\def\href#1#2{{#2}}

%%%%%%%%%%%%%%%%%%%%%%%%%%%%%%%%%%%%%%%%%%%%%%%%%%%%%%%%%%%%%%%%%
%                                                               %
%       FORMAT:                                                 %
%                                                               %
%%%%%%%%%%%%%%%%%%%%%%%%%%%%%%%%%%%%%%%%%%%%%%%%%%%%%%%%%%%%%%%%%
\parskip=3.5pt plus .3pt minus .3pt
\baselineskip=14pt plus .1pt minus .05pt
\lineskip=.5pt plus .05pt minus .05pt
\lineskiplimit=.5pt
\abovedisplayskip=18pt plus 4pt minus 2pt
\belowdisplayskip=\abovedisplayskip
\hsize=14cm
\vsize=19cm
\hoffset=1.5cm
\voffset=1.8cm
\frenchspacing
\footline={}
\raggedbottom
%%%%%%%%%%%%%%%%%%%%%%%%%%%%%%%%%%%%%%%%%%%%%%%%%%%%%%%%%%%%%%%%%
%                                                               %
%       VARIOUS DEFINITIONS                                     %
%                                                               %
%%%%%%%%%%%%%%%%%%%%%%%%%%%%%%%%%%%%%%%%%%%%%%%%%%%%%%%%%%%%%%%%%

\def\sss{\scriptscriptstyle}
\def\*{\partial}
\def\punkt{\,\,.}
\def\komma{\,\,,}

\def\={\!=\!}
\def\small#1{{\hbox{$#1$}}}
\def\half{\small{1\over2}}
\def\fraction#1{\small{1\over#1}}
\def\fr{\fraction}
\def\Fraction#1#2{\small{#1\over#2}}
\def\Fr{\Fraction}
\def\tr{\hbox{\rm tr}\,}
\def\eg{{\tenit e.g.}}

\def\a{\alpha}
\def\b{\beta}

\def\d{\delta}
\def\e{\varepsilon}
\def\g{\gamma}
\def\l{\lambda}
\def\o{\omega}

\def\O{\Omega}

\def\w{\wedge}
\def\id{1\hskip-3.5pt 1}

\def\Re{\hbox{Re}\,}
\def\Im{\hbox{Im}\,}

%%%%%%%%%%%%%%%%%%%%%%%%%%%%%%%%%%%%%%%%%%%%%%%%%%%%%%%%%%%%%%%%%%%%%%%%%%%%

\def\O{\Omega}

\def\s{\sigma}

\def\k{\kappa}

\def\sH{{\scr H}}
\def\sC{{\scr C}}
\def\sA{{\scr A}}
\def\sF{{\scr F}}

\def\Vol{\hbox{Vol}}

%%%%%%%%%%%%%%%%%%%%%%%%%%%%%%%%%%%%%%%%%%%%%%%%%%%%%%%%%%%%%%%%%%%%%%%%%%%%%

\headtext={Bao, Cederwall, Nilsson: ``A Note on
Topological M5-Branes...''}

\line{
\epsfysize=20mm
\epsffile{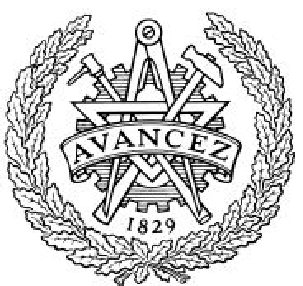}
\hfill}
\vskip-17mm
\line{\hfill G\"oteborg preprint}
\line{\hfill hep-th/0603120}
\line{\hfill March, {\old2006}}
\line{\hrulefill}

\vfill

\centerline{\sixteenhelvbold A Note on Topological M5-Branes}
\vskip6pt
\centerline{\sixteenhelvbold and String--Fivebrane Duality}

\vfill

\centerline{\twelvehelvbold Ling Bao, Martin Cederwall and Bengt E.W. Nilsson}

\vfill

\centerline{\it Fundamental Physics}
\centerline{\it Chalmers University of Technology }
\centerline{\it S-412 96 G\"oteborg, Sweden}

\vfill

{\narrower\noindent
\underbar{Abstract:}
We derive the stability conditions for the M5-brane in topological
M-theory using $\kappa$-symmetry. The non-linearly self-dual 3-form on the
world-volume is necessarily non-vanishing, as is the case also
for the 2-form field strengths on
coisotropic branes in topological string theory. It is demonstrated
that the self-duality is consistent with the stability
conditions, which are solved locally in terms of a tensor in the
representation {\bf6} of $SU(3)\subset G_2$. The double
dimensional reduction of the M5-brane is the D4-brane, and its direct
reduction is an NS5-brane. We show that the equation of motion for
the 3-form on the NS5-brane wrapping a Calabi--Yau space is exactly the
Kodaira--Spencer equation, providing support for a
string--fivebrane duality in topological string theory.
\smallskip}
\vfill

\font\xxtt=cmtt6
\def\sH{{\scr H}}
\def\sC{{\scr C}}

\vtop{\baselineskip=.6\baselineskip\xxtt
\line{\hrulefill}
\catcode`\@=11
\line{email: ling.bao@fy.chalmers.se,
martin.cederwall@chalmers.se, tfebn@fy.chalmers.se\hfill}
\catcode`\@=\active
}

\eject

\ref\CederwallNilssonSundell{M. Cederwall, B.E.W. Nilsson and P. Sundell,
    {\xit ``An action for the 5-brane in D=11 supergravity''},
    \jhep{98}{04}{1998}{007} [\hepth{9712059}].}

\ref\BaoBengtssonCederwallNilssonT{L. Bao, V. Bengtsson, M. Cederwall
    and B.E.W. Nilsson, {\xit ``Membranes for topological
    M-theory''},
    \jhep{06}{01}{2006}{150} [\hepth{0507077}].}

\ref\deBoer{J. de Boer, A. Naqvi and A. Shomer,
 {\xit ``The topological G$\sss{}_2$ string''},
        \hepth{0506211}.}

\ref\DolanNappi{L. Dolan and K. Nappi,
{\xit ``A modular invariant partition function for the five-brane''},
\NPB{530}{1998}{683} [\hepth{9806016}].}

\ref\HenningsonNilssonSalomonson{M. Henningson,
B.E.W. Nilsson and P. Salomonson,
{\xit ``Holomorphic factorization of correlation functions in
(4k+2)-dimensional (2k)-form gauge theory''},
\jhep{99}{09}{1999}{007} [\hepth{9908107}].}

\ref\WittenMfive{E. Witten, {\xit ``Five-brane effective action in M-theory''},
\JGP{22}{1997}{103} [\hepth{9610234}].}

\ref\MMMS{M. Mari\~no, R. Minasian, G.W. Moore and A. Strominger,
    {\xit ``Non-linear instantons from supersymmetric p-branes''},
        \jhep{00}{01}{2000}{005} [\hepth{9911206}].}

\ref\KapustinLi{A. Kapustin and Y. Li, {\xit ``Stability conditions
        for topological D-branes: a world sheet approach''}, \hepth{0311101}.}

\ref\GerasimovShatashvili{A.A. Gerasimov and S.L. Shatashvili, {\xit
``Towards integrability of topological strings I: Three-forms on
Calabi--Yau manifolds''}, \jhep{04}{11}{2004}{074} [\hepth{0409238}].}

\ref\DijkgraafVerlindeVonk{R. Dijkgraaf, E. Verlinde and M. Vonk,
{\xit ``On the partition sum of the NS five-brane''},
\hepth{0205281}.}

\ref\NeitzkeVafa{A. Neitzke and C. Vafa, {\xit ``N=2 strings and the
twistorial Calabi--Yau''}, \hepth{0402128}.}

\ref\NekrasovOoguriVafa{N. Nekrasov, H. Ooguri and C. Vafa, {\xit
``S-duality and topological strings''}, \jhep{04}{10}{2004}{009}
[\hepth{0403167}].}

\ref\BergshoeffLondonTownsend{E. Bergshoeff, L.A.J. London and
P.K. Townsend, {\xit ``Space-time scale invariance and the
super-p-brane''}, \CQG{9}{1992}{2545} [\hepth{9206026}].}

\ref\CederwallTownsend{M. Cederwall and P.K. Townsend, {\xit ``The manifestly
    SL(2;Z)-covariant superstring''},
    \jhep{97}{09}{1997}{003} [\hepth{9709002}]}

\ref\CederwallWesterberg{M. Cederwall and A. Westerberg,
    {\xit ``World-volume fields, SL(2;Z) and
        duality: the type IIB 3-brane''},
        \jhep{98}{02}{1998}{004} [\hepth{9710007}].}

\ref\BengtssonCederwallLarssonNilssonU{V. Bengtsson, M. Cederwall,
H. Larsson and B.E.W. Nilsson,
        {\xit ``U-duality covariant membranes''},
        \jhep{05}{02}{2005}{020} [\hepth{0406223}].}

\ref\Dijkgraafetal{R. Dijkgraaf, S. Gukov, A. Neitzke and C. Vafa,
{\xit ``Topological M-theory as unification of form theories of
gravity''}, \hepth{0411073}.}

\ref\WittenTopsigma{E. Witten, {\xit ``Topological sigma models''},
\CMP{118}{1988}{411}.}

\ref\WittenMirror{E. Witten,
{\xit ``Mirror manifolds and topological field theory''},
\hepth{9112056}.}

\ref\NeitzkeVafaReview{A. Neitzke and C. Vafa,
{\xit ``Topological strings and their physical applications''},
\hepth{0410178}.}

\ref\KodairaSpencergravity{M. Bershadsky, S. Cecotti, H. Ooguri, and C. Vafa,
{\xit ``Kodaira--Spencer theory of gravity and exact results for
quantum string amplitudes''},
\CMP{165}{1994}{311}, [\hepth{9309140}].}

\ref\Kahlergravity{M. Bershadsky and V. Sadov,
{\xit ``Theory of K\"ahler gravity''}, \IJMPA{11}{1996}{4689}
\hfill\break[\hepth{9410011}].}

\ref\BonelliZabzine{G. Bonelli and M. Zabzine, {\xit ``From current
algebras for p-branes to topological
M-theory''}, \jhep{05}{09}{2005}{015}
[\hepth{0507051}].}

\ref\Forthcoming{V. Bengtsson, M. Cederwall and B.E.W. Nilsson,
{\xit to appear.}}

\ref\OoguriOzYin{H. Ooguri, Y. Oz and Z. Yin, {\xit ``D-branes on
Calabi--Yau spaces and their mirrors''}, \NPB{477}{1996}{407}
[\hepth{9606112}].}

\ref\KapustinOrlov{A. Kapustin and D. Orlov, {\xit ``Remarks on
A-branes, mirror symmetry, and the Fukaya category''},
\hepth{0109098}.}

\ref\Chiantese{S. Chiantese, {\xit ``Isotropic A-branes and the
stability condition''}, \jhep{05}{02}{2005}{003} [\hepth{0412181}].}

\ref\GutowskiPapadopoulosTownsend{J. Gutowski, G. Papadopoulos and
P.K. Townsend, {\xit ``Supersymmetry and generalised calibrations''},
\PRD{60}{1999}{106006} [\hepth{9905156}].}

\ref\BeckerBeckerStrominger{K. Becker, M. Becker and A. Strominger,
{\xit ``Five-branes, membranes and nonperturbative string theory''},
\NPB{456}{1995}{130} [\hepth{9507158}].}

\ref\Koerber{P. Koerber, {\xit ``Stable D-branes, calibrations and
generalized Calabi--Yau geometry''}, \jhep{05}{08}{2005}{099}
[\hepth{0506154}].}

\ref\MartucciSmyth{L. Martucci and P. Smyth, {\xit ``Supersymmetric
D-branes and calibrations on general N=1 backgrounds''},
\jhep{05}{11}{2005}{048} [\hepth{0507099}].}

\ref\AnguelovaMedeirosSinkovics{L. Anguelova, P. de Medeiros and
A. Sinkovics, {\xit ``Topological membrane theory from Mathai--Quillen
formalism''}, \hepth{0507089}.}

\ref\Vonk{M. Vonk, {\xit ``A mini-course on topological strings''},
\hepth{0504147}.}

\ref\BandelloniLazzarini{G. Bandelloni and S. Lazzarini, {\xit
``Kodaira--Spencer deformation of complex structures and lagrangian
field theory''}, \JMP{39}{1998}{3619} [\hepth{9802086}].}

\ref\DuffLu{M.J. Duff and J.X. Lu, {\xit ``A duality between strings
and five-branes''}, \CQG9{1992}1.}

\ref\KapustinS{A. Kapustin, {\xit ``Gauge theory, topological strings
and S-duality''}, \jhep{04}{09}{2004}{034} [\hepth{0404041}].}

\ref\HoweSezginWest{P.S. Howe, E. Sezgin and P.C. West, {\xit
``Covariant field equations of the M theory
five-brane''}, \PLB{399}{1997}{49} [\hepth{9702008}].}

\ref\SezginSundell{E. Sezgin and P. Sundell {\xit ``Aspects of the
M5-brane''}, \hepth{9902171}.}

\ref\Nekrasov{N. Nekrasov, {\xit ``A la recherche de la m-th\'eorie
 perdue --- Z-theory: chasing m/f theory''}, %talk presented at Strings 2004,
 \hepth{0412021}.}

%%%%%%%%%%%%%%%%%%%%%%%%%%%%%%%%%%%%%%%%%%%%%%%%%%%%%%%%%%%%%%%%%%%%%%

\section\IntroAndConcl{Introduction and Conclusions}The
purpose of this paper is to formulate and examine the stability
conditions (generalised calibration relations) for M5-branes in the
topological M-theory formulated in ref. [\Dijkgraafetal] (see also
refs. [\GerasimovShatashvili,\Nekrasov]).  The stability conditions,
which have been discussed from a world sheet point of view for
D-branes in string theory in ref. [\OoguriOzYin] and for topological
string theory in refs. [\KapustinOrlov,\KapustinLi,\Chiantese], can
also be seen as a direct consequence of calibration
[\GutowskiPapadopoulosTownsend] or demanding supersymmetry
[\BeckerBeckerStrominger,\MMMS,\Koerber,\MartucciSmyth].  As is
the case \eg\ for the D4-brane in the A-model, the stability conditions demand
non-vanishing world-volume field strength. Here we derive the
corresponding stability conditions for the M5-brane in topological
M-theory and its close relative the NS5-brane in the topological
A-model. This is achieved using the $\k$-symmetric top-form
formulation applied to the physical M5-brane in
ref. [\CederwallNilssonSundell].  In this approach there is in the
7-dimensional $G_2$ superspace, apart from the super-4-form field
strength, also a super-7-form field strength obeying the appropriate
Bianchi identities, but without a bosonic component.

The M5-brane is, apart from the topological membrane constructed in
ref. [\BaoBengtssonCederwallNilssonT] (for a different approach see
ref. [\BonelliZabzine,\AnguelovaMedeirosSinkovics]), the only brane
present in topological M-theory\foot\dagger{{\xit G}$\sss{}_2$ target
spaces occur
also in the topological string constructed in ref. [\deBoer]; its
relation to topological M-theory is, however, unclear to us.}. Their
direct and double dimensional reductions on a circle to a Calabi--Yau
space give all NS-branes and D-branes in the A-model save for the
isotropic D-branes with one-dimensional world sheets introduced in
ref. [\Chiantese] which should probably be viewed as Kaluza--Klein
modes.

We proceed to demonstrate how the direct reduction of the M5-brane on
CY$\times S^1$ gives the NS5-brane in the topological A-model
introduced in ref. [\WittenTopsigma,\WittenMirror] (see
refs. [\NeitzkeVafa,\Vonk] for a review of topological string theory),
whose world-volume inherits the dynamical Kodaira--Spencer deformation
theory [\BandelloniLazzarini] from the M5-brane.  The related
connection between the M5-brane instantons in the physical M-theory
and Kodaira--Spencer theory was first pointed out in ref. [\MMMS].
The double reduction will give the D4-brane, with the stability
conditions formulated by Kapustin and Li [\KapustinLi] (although that
correspondence is not shown in the present paper). The NS5-brane
provides a precise description of how duality between K\"ahler gravity
[\Kahlergravity] and Kodaira--Spencer theory [\KodairaSpencergravity],
describing deformations of the K\"ahler and complex structures,
respectively, is realised in the A-model as a ``string--fivebrane
duality'' [\DuffLu]. A forthcoming paper [\Forthcoming] will extend
the discussion to the full sets of D-branes and RR fields in the A-
and B-models.

Related conjectures have been made earlier. In
ref. [\DijkgraafVerlindeVonk] Dijkgraaf, Verlinde and Vonk used
T-duality to relate the partition function on coinciding NS5-branes
(with linear self-duality) in the A-model to a B-model calculation.
S-duality, relating the A- and B-models on the same manifold, for
topological strings, was conjectured on a twistorial CY by Neitzke and
Vafa [\NeitzkeVafa], and clarified, mainly using D-instantons, by
Nekrasov, Ooguri and Vafa in ref. [\NekrasovOoguriVafa], where the
existence of the topological NS5-brane was also pointed out. The
relevance of the calculation of ref. [\DijkgraafVerlindeVonk] in this
context was observed in ref. [\KapustinS].  Gerasimov and Shatashvili,
in their paper pointing towards a topological M-theory
[\GerasimovShatashvili], relate Kodaira--Spencer theory to a
7-dimensional theory. Mari\~no et al.  [\MMMS] derive conditions for
$D=11$ M5-branes wrapping a Calabi--Yau space to preserve
supersymmetry, and derive the Kodaira--Spencer equation. We comment to
the relation of the present paper to the latter work in section 3.

\section\TopMFiveBranes{Topological M5-branes}The
reduction of topological M-theory on a circle contains the
A-model [\Dijkgraafetal].
The presence in the A-model of a D4-brane and an NS5-brane
implies that there has to be a 5-brane in topological M-theory. The
purpose of this section is to derive, using superspace techniques and
$\k$-symmetry, the stability conditions for this topological M5-brane,
and to demonstrate the consistency between these conditions and the
non-linear self-duality for the 3-form field strength on the
brane. Open topological membranes have boundaries on the 5-brane, just
as fundamental strings end on D-branes and D-branes on NS5-branes in
the A-model.

We work in a superspace with 16 real fermionic directions and
R-symmetry group $SL(2)$. This is half the number of fermionic
coordinates compared to superstring theory or M-theory, and
appropriate for the formulation of a topological 7-or 6-dimensional
theory\foot\dagger{A more systematic formulation of topological
theories embedded in supergravities with 16 supercharges will be given
in a forthcoming paper [\Forthcoming].}.  The dimension-0 components
of the torsion and the 4-form field strength are
$$
\eqalign{
T^a_{\a I,\b J}&=2\e_{IJ}\g^a_{\a\b}\komma\cr
H_{ab,\a I,\b J}&=2\e_{IJ}(\g_{ab})_{\a\b}\punkt\cr
}\eqn
$$
The real $\g$-matrices, which can be viewed as imaginary unit
octonions multiplying octonionic spinors of $Spin(7)$, square to $-1$.
For details on $\gamma$-matrices etc., we refer to the Appendix and to
ref. [\BaoBengtssonCederwallNilssonT].

Even though there is no bosonic 7-form field strength in the
supergravity multiplet, there is a 7-form field strength on
superspace, namely
$$
\sH_{abcde,\a I,\b J}=2\e_{IJ}(\g_{abcde})_{\a\b}\komma\eqn
$$
with the Bianchi identity $d\sH+\half H\w H=0$, following from the
7-dimensional Fierz identities. This presence of a superspace field
strength that does not contain a purely bosonic part, or, more
precisely, the absence of an invariant cohomologically non-trivial
6-form to calibrate the 6-cycle of the brane world-volume, is symptomatic
for the cases of high-dimensional branes where non-vanishing
world-volume field strength is demanded by the generalised calibration
(stability) conditions.

We write an action for the 5-brane in complete analogy with
ref. [\CederwallNilssonSundell], the only difference being that the
signature of the world-volume is euclidean,
$$
S=\int d^6\xi\sqrt{g}\l\left[1+\Phi(F)+(\star\sF)^2\right]\komma
\Eqn\FivebraneAction
$$
where the field $\l$ is a Lagrange multiplier and $\Phi$ a functional
to be determined. $\sF$ is the modified 6-form field strength of a
5-form potential $\sA$ and the 3-form $F$ is the field strength of the
2-form $A$:
$$
\eqalign{
F&=dA-C\komma\cr
\sF&=d\sA-\sC-\half A\w H\cr
}\eqn
$$
where the pullbacked superfield potentials $C$ and ${\sC}$ provide
the coupling to the background. These field strengths are
constructed with background gauge invariance as guideline. The
Bianchi identities are $dF=-H$, $d\sF=-\sH+\half F\w H$.  The action
has of course to be supplemented by some self-duality condition. The
advantage of actions of this type
[\BergshoeffLondonTownsend,\CederwallTownsend,\CederwallWesterberg,\CederwallNilssonSundell,\BengtssonCederwallLarssonNilssonU],
with world-volume fields corresponding to all background fields the
brane couples to, is (apart from complete control over background
couplings and possible boundary conditions for lower-dimensional
branes) that consistency of the non-linear self-duality relation is
restrictive enough that demanding $\k$-symmetry gives its explicit
form, which can be obtained without {\it a priori} specifying the
function $\Phi$. At the same time, the corresponding projector on
$\k$ is derived, and $\Phi$ can be constructed.

We define $K^{ijk}\equiv{\partial\Phi\over\partial F_{ijk}}$. The
equations of motion for $A$, $\sA$ and $\lambda$ are
$$
\eqalign{
&d(\l{\star K})-\l(\star\sF)H=0\komma\cr
&d(\l{\star\sF})=0\komma\cr
&1+\Phi+(\star\sF)^2=0\komma\cr
}\eqn
$$
respectively. These must be consistent with the Bianchi
identities, thus, combining the first two equations of motion with
the Bianchi identity $dF=-H$ we find $K=(\star\sF)\star F$. By
varying the action using $\delta_\k F=-i_\k H$ and
$\delta_\k\sF=-i_\k\sH+\half F\w i_\k H$ and inserting the
relation between $K$ and the field strengths, the projection
matrix on $\k$ and the non-linear self-duality of the field
strengths are obtained. We leave out the details, since they are
in close parallel to ref. [\CederwallNilssonSundell], and state
the result. For the action to be invariant under the $\k$-symmetry
the parameter $\k$ must satisfy $(1-\Gamma)\k=0$, with
$$
\Gamma={i\over N\sqrt{g}}\e^{ijklmn}\left[\fr{6!}\g_{ijklmn}
+\fr{2(3!)^2}F_{ijk}\g_{lmn}\right]\Eqn\GammaMatrix
$$
and $N\equiv\sqrt{1+\Phi}$. The self-duality relation is
$$
iN{\star F}_{ijk}=N^2F_{ijk}+\half q_{[i}{}^lF_{jk]l}\komma\Eqn\SelfDuality
$$
where the sign choice $\star\sF=-i\sqrt{1+\Phi}=-iN$ has been
used. Here we have introduced the symmetric matrix $k_{ij}=\half
F_i{}^{kl}F_{jkl}$ and the traceless $q=k-\fr6\tr k$. Inserting
eq. (\SelfDuality), together with the Bianchi identities, into the
equations of motion we find $\Phi=-\fr6\tr k-\fr{24}\tr
q^2+\fr{144}(\tr k)^2$. On the other hand, contracting the
self-duality relation (\SelfDuality) with $F^{ijk}$ gives $\tr
q^2=-24N^2(1-N^2)$, which by representation theory turns out to be the
stronger relation $q^2=-4N^2(1-N^2)\id$. The equation of motion for
the Lagrange multiplier $\lambda$ now becomes
$$
N^2=1-\fr{12}\tr k\punkt\eqn
$$
This relation follows in fact also from $\Gamma^2=\id$. Dualising
the self-duality relation (and using all the known relations between
$N$, $k$ and $q$ as well as ``$\star(qF)=-q{\star F}$'') gives
consistency.

After elimination of the top-form $\sF$, we may write an action of a
more standard type giving the same equations of motion,
$$
S=\int d^6\xi\sqrt{g(1+\Phi)}+i\int(\sC-\half F\w C)\punkt\eqn
$$
Although this type of action (supplemented with some
self-duality\foot\dagger{Note that the implementation of the
self-duality condition [\CederwallNilssonSundell]
can only be done on the level of the partition
function, see [\WittenMfive, \DolanNappi,
\HenningsonNilssonSalomonson].}) is less convenient as a starting
point, the calibration relations we derive below has a clearer
interpretation as relating kinetic and Wess--Zumino terms, as usual.

We are now ready to consider this M5-brane in a manifold with $G_2$
holonomy, and look for 6-cycles that, together with the appropriate
values of $F$, preserve supersymmetry. There is a covariantly constant
spinor $\eta^I$ (for each value of the $SL(2)$ index $I$), which we
take to be the real part of the octonion. We expect the global
supersymmetry to play the r\^ole of BRST charges, in analogy with the
situation for the topological membrane of
ref. [\BaoBengtssonCederwallNilssonT]. Using the the methods of that
reference it should be possible that the action (\FivebraneAction) is
not only BRST-invariant (supersymmetric) but also BRST-exact.

Using the explicit
expressions for $\g$-matrices in terms of $G_2$-invariant tensors we
have the action of $\Gamma$ on the covariantly constant spinor:
$$
\Gamma\left[\matrix{1\cr0\cr}\right]={i\over N\sqrt{g}}\e^{ijklmn}
\left[\matrix{\fr{2(3!)^2}F_{ijk}\s_{lmn}\cr
-\fr{6!\sqrt{g}}\e_{ijklmn}\d^\a_7+\fr{2(3!)^2}F_{ijk}{\star\s}_{lmn\a}\cr}\right]
\komma\eqn\
$$
where we for convenience have used a local basis where the direction
$dx^7$ is normal to the world-volume. The tensor $\s$ is the
covariantly constant $G_2$-invariant 3-form. The criterion for
supersymmetry is that $(1-\Gamma)\eta=0$, which yields the stability
conditions for the brane:
$$
\eqalign{
&\Fr i2F\w f^*\s=N\Vol_6\komma\cr
&\fr2F\w{\star\s}=-\Vol_7\komma\cr
&F\w i_vf^*{\star\s}=0\komma\cr
}\Eqn\FirstStabCond
$$
where $\Vol_6$ and $\Vol_7$ are the world-volume and space volume
forms, respectively, and $v$ is any world-volume vector. In order to
solve these relations locally, and check their consistency, we
parametrise the tensors using the local breaking of $G_2$ to
$SU(3)$, and use the standard relations $\s=\Re\O+\o\w dx^7$,
$\star\s=-\Im\O\w dx^7-\half\o\w\o$ (see appendix for conventions).
At the moment this is not necessarily to be seen as the direct
reduction to an A-model NS5-brane, although the local
parametrisation suits this case. The $SU(3)$-covariant version of
the stability conditions is
$$
\eqalign{
&\Fr i2F\w f^*\Re\O=N\Vol_6\komma\cr
&\fr2F\w f^*\Im\O=\Vol_6\komma\cr
&F\w f^*\o=0\punkt\cr
}\Eqn\StabilityConditions
$$
From the conditions (\StabilityConditions) it follows immediately
that $F_{abc}=-\Fr{1+N}4\O_{abc}$, $F_{\bar a\bar b\bar
c}=-\Fr{1-N}4\bar\O_{\bar a\bar b\bar c}$ (we suppress explicit
pullbacks from now on), and that $g^{b\bar c}F_{ab\bar c}=0$ and
$g^{a\bar b}F_{a\bar b\bar c}=0$ (the last two equations leave only
the representations ${\bf\bar6}$ out of ${\bf\bar6}\oplus{\bf3}$ in
$F_{(2,1)}$ and ${\bf6}$ out of ${\bf6}\oplus{\bf\bar3}$ in
$F_{(1,2)}$). It is not {\it a priori} clear that the stability
conditions, derived from the $G_2$ structure, are consistent with
the self-duality relations. We will however show that this is indeed
the case, and that, given the value of $F_{(3,0)}$ from the
stability condition, the self-duality relation dictates exactly the
value of $F_{(0,3)}$ given after eq. (\StabilityConditions).

It is convenient to parametrise the non-linearly self-dual 3-form $F$
in terms of a linearly self-dual one, $h$. It is straightforward to
show that $h_{ijk}=F_{ijk}+\fr{2N(1+N)}q_i{}^lF_{jkl}$ satisfies
$i{\star h}=h$. Forming the matrix $r_{ij}=\half h_{i}{}^{kl}h_{jkl}$,
the relations above give $r=\Fr2{N(1+N)}q$, so the relation between
$h$ and $F$ becomes $h_{ijk}=m_i{}^lF_{jkl}$, where
$m=\id+\fr4r$. Inverting the matrix $m$,
$m^{-1}=\Fr{(1+N)}2(\id-\fr4r)$, finally gives the explicit
parametrisation of $F$ in terms of $h$,
$$
F_{ijk}=\Fr{1+N}2\left(h_{ijk}-\fr4r_i{}^lh_{jkl}\right)\komma\eqn
$$
where the scalar $N$ now is defined by $r^2=-16{1-N\over1+N}\id$.

The general Ansatz for $h$ in terms of $SU(3)$ tensors contains a
singlet in $h_{(3,0)}$ ($\xi$), a triplet ${\bf3}$ in $h_{(2,1)}$ and the
representation ${\bf6}$ in $h_{(1,2)}$ ($u$). It is clear that the triplet
generates triplets in $F$ violating the last equation in
(\StabilityConditions), so we set it to zero. The Ansatz becomes
$$
\eqalign{
h_{abc}&=\half\xi\O_{abc}\komma\cr
h_{ab\bar c}&=0\komma\cr
h_{a\bar b\bar c}
         &=\fr2 u_a{}^{\bar d}\bar\O_{\bar b\bar c\bar d}\komma\cr
h_{\bar a\bar b\bar c}&=0\punkt\cr
}\eqn
$$
The matrix $r$ has the non-vanishing components $r_{ab}=4\xi
u_a{}^{\bar c}g_{b\bar c}$, $r_{\bar a\bar b}=\fr4\bar\O_{\bar
a}{}^{cd} \bar\O_{\bar b\bar e\bar f}u_c{}^{\bar e}u_d{}^{\bar f}$.
Calculating $F$ from this Ansatz gives immediately
$F_{abc}=\Fr{1+N}4\xi\O_{abc}$, so $\xi=1$ by the stability
conditions. We have $\tr r^2=96\det u$ (note that $\det
u=\fr{8\cdot3!}\bar\O^{abc}\bar\O_{\bar a\bar b\bar c} u_a{}^{\bar
a}u_b{}^{\bar b}u_c{}^{\bar c}$), and thus $\det u={1-N\over1+N}$.
The complete non-linearly self-dual tensor is
$$
\eqalign{
F_{abc}&=-\Fr{1+N}4\O_{abc}\komma\cr
F_{ab\bar c}&=\Fr{1+N}4\bar\O_{\bar c\bar d\bar e}u_a{}^{\bar
d}u_b{}^{\bar e}=\Fr{1-N}4\O_{abd}(u^{-1})_{\bar c}{}^d\komma\cr
F_{a\bar b\bar c}
         &=\Fr{1+N}4 u_a{}^{\bar d}\bar\O_{\bar b\bar c\bar d}\komma\cr
F_{\bar a\bar b\bar c}&=-\Fr{1+N}4\bar\O_{\bar a\bar b\bar c}\det u
        =-\Fr{1-N}4\bar\O_{\bar a\bar b\bar c}\punkt\cr
}\eqn
$$
We notice that the value of $F_{(0,3)}$ consistent with the stability
conditions is exactly the one that follows from non-linear
self-duality. This concludes the check of algebraic consistency of the
stability conditions (\FirstStabCond) with the self-duality relation
(\SelfDuality), and provides an explicit parametrisation for the
following section.

%\vfill\eject

\section\NSBraneKS{NS5-Branes in the A-model and Kodaira--Spencer
Theory}So
far, the analysis is completely local and algebraic. We will show
that the equation of motion (or equivalently, the Bianchi identity)
for the 3-form is the Kodaira--Spencer equation. We will now suppose
that the M5-brane actually winds a Calabi--Yau space, so that it
becomes an NS5-brane in the A-model. The components of $dF=0$ are (we
assume that the RR field strengths vanish)
$$
\eqalign{
(dF)_{(1,3)}:\quad
&\partial_aN-\bar\partial_{\bar b}[(1+N)u_a{}^{\bar b}]=0\komma\cr
(dF)_{(2,2)}:\quad
&\bar\O^{acd}\partial_c[(1+N)u_d{}^{\bar b}]
     +\O^{\bar b\bar c\bar d}
     \bar\partial_{\bar c}[(1-N)(u^{-1})_{\bar d}{}^a]=0\komma\cr
(dF)_{(3,1)}:\quad
&\bar\partial_{\bar a}N+\partial_b[(1-N)(u^{-1})_{\bar a}{}^b]=0\punkt\cr
}\Eqn\dFisZero
$$
It is straightforward to show, using $dN=-\half(1-N^2)\tr(u^{-1}du)$,
that the first two equations imply the third. The first equation can
be seen as a gauge-fixing condition, while the second one reads
$$
\eqalign{
0&=\partial_{[a}[(1+N)u_{b]}{}^{\bar c}]
+\bar\partial_{\bar d}[(1+N)u_{[a}{}^{\bar d}u_{b]}{}^{\bar c}]\cr
&=\bigl(\partial_{[a}N
-\bar\partial_{\bar d}[(1+N)u_{[a}{}^{\bar d}]\bigr)u_{b]}{}^{\bar c}
+(1+N)\bigl(\partial_{[a}u_{b]}{}^{\bar c}
   -u_{[a}{}^{\bar d}\bar\partial_{\bar d}u_{b]}{}^{\bar c}\bigr)\komma
}\eqn
$$
which, using the gauge-fixing condition, implies that $u$ fulfills the
Kodaira--Spencer equation
$$
\partial_{[a}u_{b]}{}^{\bar c}
   -u_{[a}{}^{\bar d}\bar\partial_{\bar d}u_{b]}{}^{\bar c}=0\komma\eqn
$$
corresponding to the deformation of the complex structure encoded in
the differential $\partial' =dz^a(\partial_a-u_a{}^{\bar
b}\bar\partial_{\bar b})$.

The non-linearly self-dual closed 3-form $F$ is exactly the
deformation of the form $\half\O$ defining the complex structure. It
will be linearly self-dual under the deformed metric. It is possible
to be quite explicit about the deformed metric $G$, such that
$i{\star_G}F=F$. From the form of the non-linear self-duality
relation, it is clear that the metric $G$ satisfies (using that the
antisymmetry of $G_{[i}{}^l F_{jk]l}$ is automatic provided $G$ is
expressible in terms of $F$)
$$
{G^3\over\sqrt{\det G}}=N\id-{1\over2N}q\komma\eqn
$$
where contractions are made with the undeformed metric (which we for
calculations have taken to be locally $\id$).  The right hand side has
unit determinant.  The expressions become more transparent if we use
the normalised matrix $s={1\over2N\sqrt{1-N^2}}q$ with $s^2=-\id$. We
then have $(\det G)^{-1/2}G^3=N\id-\sqrt{1-N^2}s=e^{-s\theta}$, where
%the angle
$\theta$ is defined by $\cos\theta=N$. The deformed metric
is thus defined, up to a scale factor, by
$$
(\det G)^{-1/6}G=e^{-{1\over3}s\theta}\punkt\eqn
$$
It will of course be hermitean only with respect to the deformed
complex structure.

We would like to comment on the relation to the treatment of the
11-dimensional M5-brane instantons winding on CY spaces of
ref. [\MMMS]. The projection matrix on the $\k$ parameter stated there
does not contain the actual $\Gamma$ of eq. (\GammaMatrix), but only
its linearisation in $h$, which is the projection arising from a
superembedding treatment [\HoweSezginWest].  It was shown in
ref. [\SezginSundell] how the two apparently different projections
``$\half(1-\Gamma)$'' are related, and that they both project on the
fermionic gauge degrees of freedom. Here we start from a topological
M5-brane, in a superspace with 7 bosonic coordinates and half the
number of fermions compared to M-theory, whose presence in topological
M-theory is necessitated by the existence of D4- and NS5-branes in the
A-model.

%\vfill\eject

\appendix{Conventions}In 7
euclidean dimensions, we use $\gamma$ matrices that satisfy
$
\{ \gamma^a,\gamma^b \} = -2\delta^{ab}\komma
$
where the minus sign is necessary for real $\g$-matrices. The
spinors are real $\psi_{\a}^I$, where $\a=1,..,8$ and the $I=1,2$ is
an $SL(2,R)$ $R$-symmetry index [\BaoBengtssonCederwallNilssonT].

For the 3-form $\s$, we use $\s_{124}=1$ and cyclic. On the CY
space, with 3 complex dimensions, we use locally
$\O_{abc}=\e_{abc}$, so that $\O\w\bar\O=8i\Vol_6$. We have
$g_{a\bar b}=\half\delta_{a\bar b}$ and $\o_{a\bar b}=\Fr
i4\delta_{a\bar b}$, so that $\o\w\o\w\o=-6\Vol_6$. The relations
between 7-dimensional and 6-dimensional forms are
$$
\eqalign{
\s&=\Re\O+\o\w dx^7\komma\cr
\star\s&=-\Im\O\w dx^7-\half\o\w\o\punkt\cr
}\eqn
$$

The real 7-dimensional $\gamma$ matrices encoded in the left
multiplication of a spinor $\lambda=\lambda^{\hat\a}e_{\hat\a}$ by an
imaginary unit $e_a$ are
$$
\eqalign{
(\g^a)_{\a\b}&=\s^a{}_{\a\b}\komma\cr
(\g^a)_{0\a}&=\d^a_\a\punkt\cr
}\eqn
$$

The Clifford algebra is spanned by the $so(7)$-invariant tensors
$\delta^{\hat{\alpha}}_{\phantom{\hat{\alpha}}\hat{\beta}}$,
$(\gamma^a)^{\hat{\alpha}}_{\phantom{\hat{\alpha}}\hat{\beta}}$,
$(\gamma^{ab})^{\hat{\alpha}}_{\phantom{\hat{\alpha}}\hat{\beta}}$ and
$(\gamma^{abc})^{\hat{\alpha}}_{\phantom{\hat{\alpha}}\hat{\beta}}$,
of which the first and last are symmetric and the second and third
antisymmetric matrices. The decomposition in terms of $G_2$-invariant
tensors is
$$
\eqalign{
\delta^{\hat{\alpha}}_{\phantom{\hat{\alpha}}\hat{\beta}} & = \left[
\matrix{ 1 & 0 \cr
0 & \delta^{\alpha}_{\phantom{\alpha}\beta} \cr
} \right] \komma\cr
(\gamma^a)^{\hat{\alpha}}_{\phantom{\hat{\alpha}}\hat{\beta}} & = \left[
\matrix{ 0 & \delta^{a}_{\phantom{a}\beta} \cr
-\delta^{a\alpha} & \sigma^{a\alpha}_{\phantom{a\alpha}\beta} \cr
} \right] \komma\cr
(\gamma^{ab})^{\hat{\alpha}}_{\phantom{\hat{\alpha}}\hat{\beta}} & = \left[
\matrix{ 0 & -\sigma^{ab}_{\phantom{ab}\beta} \cr
\sigma^{ab\alpha} & -\star\sigma^{ab\alpha}_{\phantom{ab\alpha}\beta}
- 2\d^{ab}_{\alpha\beta} \cr
} \right] \komma\cr
(\gamma^{abc})^{\hat{\alpha}}_{\phantom{\hat{\alpha}}\hat{\beta}} & = \left[
\matrix{ \sigma^{abc} & -\star\sigma^{abc}_{\phantom{abc}\beta} \cr
-\star\sigma^{abc\alpha} & 6
\delta_{(\alpha}^{[a}\sigma_{\beta)}^{\phantom{\beta)}bc]} -
\delta^{\alpha}_{\phantom{\alpha}\beta}\sigma^{abc} \cr }
\right]\punkt
}\eqn
$$

%\vfill\eject

\refout
\end